\documentstyle[multicol,prl,aps,epsf]{revtex}

\begin{document}
\title{Quantum interference and supercurrent
in multiple-barrier proximity structures}
\author{ Artem V. Galaktionov and Andrei D. Zaikin}
\address{Forshchungszentrum Karlsruhe, Institut f\"ur Nanotechnologie,
76021, Karlsruhe, Germany\\
I.E. Tamm Department of Theoretical Physics, P.N. Lebedev Physics
Institute, Leninskii pr. 53, 119991 Moscow, Russia}
\maketitle

\begin{abstract}
We analyze an interplay between the proximity effect and quantum interference
of electrons in hybrid structures superconductor-normal metal-superconductor
which contain several insulating barriers. We demonstrate that the dc
Josephson current in these structures may change qualitatively due to quantum
interference of electrons scattered at different interfaces. In junctions with
few conducting channels mesoscopic fluctuations of the supercurrent are
significant and its amplitude can be strongly enhanced due to resonant
effects. In the many channel limit averaging over the scattering phase
effectively suppresses interference effects for systems with {\it two}
insulating barriers. In that case a standard quasiclassical approach
describing scattering at interfaces by means of Zaitsev boundary conditions
allows to reproduce the correct results. However, in systems with {\it three}
or more barriers the latter approach fails even in the many channel limit. In
such systems interference effects remain important in this limit as well. For
short junctions these effects result in additional suppression of the
Josephson critical current indicating the tendency of the system towards
localization. For relatively long junctions interference effects may -- on the
contrary -- enhance the supercurrent with respect to the case of independent
barriers.

\end{abstract}

\begin{multicols}{2}

\section{Introduction}
In recent years there has been a great deal of activity devoted to
both experimental and theoretical studies of mesoscopic superconducting-normal
($SN$) hybrid structures\cite{lam,bel}). New
important phenomena such as anomalous Meissner screening,
re-entrant behavior of the
conductance, nonequilibrium-driven $\pi$-junction state and many others have
been discovered and thoroughly investigated. In some cases it was
found that an interplay between the proximity effect and
quantum interference of electrons in a normal metal play a significant
role at sufficiently low temperatures. For instance, interference of
electrons scattered at impurities in the normal layer may
strongly enhance the Andreev conductance $G_{A}$ of $SN$
systems leading to the so-called zero-bias anomaly\cite{VZK,HN,B}:
At low voltages
and temperatures $G_A$ turns out to depend linearly on the
$SN$ interface transmission $D$ in contrast to the standard result
$G_A\propto D^2$ obtained in the absence of interference effects
in the $N$-layer.

In this paper we will address a different -- although somewhat related --
problem. We will analyze the dc Josephson effect in $SNS$ systems which
contain several insulating barriers. In this case electrons scattered at
different barriers can interfere inside the junction. We will demonstrate that
this effect may lead to qualitative modifications of the supercurrent across
the junction. The most pronounced effect of quantum interference is expected
in $SNS$ systems with few conducting channels. This situation can be realized,
for instance, if two superconductors are connected via a carbon nanotube
\cite{Kas,Christ}. More conventional $SNS$ structures with many
conducting channels and several insulating barriers are also of
considerable interest, for instance in relation to possible
applications, see e.g. Ref. \onlinecite{Misha} and further references
therein. We will demonstrate that for such systems
quantum interference effects are also important
provided there exist more than two scatterers inside the junction.

A powerful tool for theoretical studies of mesoscopic
superconductivity is provided by the quasiclassical formalism of
the energy-integrated Eilenberger Green functions \cite{Eil}
(see also Refs. \onlinecite{lam,bel,lar,Albert} for a review).
The Eilenberger equations are, on one hand, much simpler than
the fully microscopic Gorkov equations and,
on the other hand, allow to correctly describe the system behavior at
distances much longer as compared to the Fermi wavelength $1/k_F$.
Since typical length scales in superconductors
(e.g. the coherence length $\xi_0$ or the London penetration
depth) are all several orders of magnitude greater than $1/k_F$, the
quasiclassical approach is usually an excellent approximation.

The quasiclassical equations cannot be applied only in the vicinity
of inter-metallic interfaces and barriers where rapid changes
of the system properties (at scales comparable to $1/k_F$) occur.
Fortunately, in many cases this problem can be circumvented by
matching the Eilenberger Green functions on both sides of the interface
with the aid of the proper boundary conditions. In order to derive
such conditions it is necessary to go beyond quasiclassics, however
under certain assumptions the final
result can be formulated
only in terms of the quasiclassical Eilenberger propagators. The derivation of
these boundary conditions was performed by Zaitsev\cite{zait}. Supplemented
by these boundary conditions, the Eilenberger quasiclassical formalism
was proven to be an extremely efficient tool for a quantitative description of
numerous inhomogeneous and hybrid superconducting structures.

An important ingredient of the derivation\cite{zait} is the
assumption that the
boundaries are situated sufficiently far from each other,
so that {\it interference effects} emerging from scattering at different
interfaces can be totally neglected. Under this assumption one arrives
at the nonlinear matching conditions involving the
third power of the quasiclassical propagators. These matching conditions
are expressed in terms of the interface transparency coefficients
for the electrons with different directions of the Fermi
momenta. It is also essential that Zaitsev boundary conditions do
not depend on scattering phases at the interface potentials.

Although for metallic structures containing one
interface one can indeed disregard interference effects,
in systems with several boundaries this is in general not
anymore possible. Hence, the applicability of
the nonlinear matching conditions\cite{zait} to multiple-barrier
systems requires additional analysis.
Some authors\cite{j1} argued that the standard quasiclassical approach
can break down in a multiple-interface geometry due to the problems with the
normalization of the Eilenberger functions.

In principle the above problem with boundary conditions can be avoided
within the approach based on the Bogolyubov-de Gennes
equations\cite{dG}. However, this approach, though frequently successful, may
also be technically inconvenient in complicated situations, for
instance because of a necessity to evaluate the energy eigenvalues
of the system and to perform summation over the energy spectrum
in the final results.

It is also possible to formulate an alternative quasiclassical
approach\cite{PZ} which allows to avoid the
abovementioned problems. Without going into details here let us just
mention that within the technique\cite{PZ} one deals with the quasiclassical
spinor amplitude $u,v$-functions which depend on
one coordinate and one time only and obey linear first order equations.
The Eilenberger Green functions are expressed via two
linearly-independent solutions of these equations
in a way that both the Eilenberger equations and
the normalization conditions are automatically satisfied. The
formalism\cite{PZ} -- just as the Eilenberger one -- can be formulated
both within the Matsubara and the Keldysh techniques and thus is suitable
both in equilibrium and non-equilibrium situations (see,
e.g. Ref. \onlinecite{GZ}) in various superconducting structures.
It is also important that very simple {\it linear} boundary conditions
for the quasiclassical amplitudes can be formulated at each of
the interfaces where electron scattering takes place. The number of
interfaces in the system is not restricted and the interference effects are
properly taken care of. Thus it is possible to take
advantage of the quasiclassical approximation and at the same time
to formulate general and simple boundary conditions without making
additional assumptions employed in Ref. \onlinecite{zait}.

Similar ideas have recently been put forward by Shelankov and
Ozana\cite{shel}. These authors also used linear matching conditions (obtained
by means of the scattering matrix approach) for the ``wave functions'' which
factorize the two-point Green functions. The next step\cite{shel} was to
construct quasiclassical one-point Green functions and formulate {\it
nonlinear} boundary conditions for such functions which would now adequately
include information about scattering on arbitrary number of ``knots''. Linear
boundary conditions were also used by Brinkman and Golubov\cite{brink} in a
calculation related to ours (see below).

In this paper, following Refs. \onlinecite{PZ,shel,brink}, we will
use simple linear boundary conditions in order to match the
quasiclassical amplitude functions at interfaces. However, unlike in
Ref. \onlinecite{shel}, we will avoid reformulating these boundary
conditions as nonlinear
ones for the Eilenberger Green functions. Rather we will directly
express the two-point Green functions and the expectation value of
the current operator in terms of the quasiclassical amplitudes.
We will then apply our method to the calculation of dc
Josephson currents in hybrid $SINI'S$ and $SINI'NI''S$ structures in
the clean limit and for arbitrary interface transmission coefficients

The interference of the scattering events at different interfaces manifests
itself in the expressions containing scattering phases $\phi$ at the interface
potentials. For the systems with two barriers (in our case
$SINI'S$-systems) with {\it many transmission channels} the
summation over their contributions is equivalent to effective
averaging over $\phi$. In this limit one can demonstrate that {\it after}
such averaging our result is equivalent one obtained from the
Eilenberger equations supplemented by the Zaitsev boundary
conditions. However, in the case of more than two barriers (i.e. for
$SINI'NI''S$ junctions) the dependence on
the scattering phases turns out to be much more essential. In this
situation the approach employing Zaitsev boundary conditions
turns out to fail also in the many channel limit where quantum interference
effects survive even after averaging over the scattering phases.

The paper is organized as follows. Our quasiclassical approach is
outlined in Sec. II. In Sec. III we apply this approach for the
analysis of the dc Josephson effect in $SINI'S$ structures with
arbitrary interface transmissions. The Josephson
current across $SINI'NI''S$ structures is evaluated in Sec. IV.
In Sec. V we present a brief discussion and summary of our results.
Some technical details of our calculation are relegated to Appendices.

\section{General method}
\subsection{Quasiclassical approximation}
The starting point of our analysis are the microscopic Gor'kov
equations\cite{AGD}. In what follows we will
assume that our system is uniform along the directions parallel to
the interfaces (coordinates $y$ and $z$).
Performing the Fourier transformation  of the normal
$G$ and anomalous $F^+$ Green function with respect to these coordinates
$$ G_{\omega_n}(\bbox{r}, \bbox{r'})=
\int\frac{d^2 \bbox{k_\parallel}}{(2\pi)^2}
G_{\omega_n} (x,x',\bbox{k_\parallel})e^{i
\bbox{k_\parallel}(\bbox{r_\parallel}- \bbox{r'_\parallel})}
$$
we express the Gor'kov equations in the following standard form
\begin{equation}
{\small \left( \begin{array}{cc} i\omega_n -\hat H & \Delta(x)\\ \Delta^*(x)&
i\omega_n +\hat H_c\end{array}\right)\left( \begin{array}{c} G_{\omega_n}
(x,x',\bbox{k_\parallel})\\F^+_{\omega_n} (x,x',\bbox{k_\parallel})
\end{array}\right)= \left(\begin{array}{c} \delta(x-x')\\ 0\end{array}
\right).} \label{start}
\end{equation}
Here $\omega_n=(2n+1)\pi T$ is the Matsubara frequency, and $\Delta(x)$ is the
superconducting order parameter. The Hamiltonian $\hat H$ in Eq.(\ref{start})
reads
\begin{equation}
\hat H=-\frac{1}{2m}\frac{\partial^2}{\partial x^2}+
\frac{\bbox{\tilde k^2_\parallel}}{2m}-\epsilon_F +V(x).
\label{H}
\end{equation}
Here $\bbox{\tilde k_\parallel}= \bbox{k_\parallel}-
\frac{e}{c}\bbox{A_\parallel}(x)$, $\epsilon_F$ is Fermi energy,
the term $V(x)$ accounts for the external
potentials (including the boundary potential). The Hamiltonian $\hat
H_c$ is obtained from $\hat H$ (\ref{H}) by inverting the sign of the electron
charge $e$. The above Hamiltonians can also include the self-energy
terms which, however, will not be considered below.

The quasiclassical approximation makes it possible to conveniently
separate fast oscillations of the Green functions due to the factor
$\exp(\pm ik_x x)$ from the
envelope of these functions changing at much longer scales as compared
to the atomic ones. Making use of this approximation for
two-component vector $\overline{\varphi}_\pm(x)\exp(\pm ik_x x)$ we
obtain
$$
\left( \begin{array}{cc} i\omega_n -\hat H & \Delta(x)\\ \Delta^*(x)&
i\omega_n +\hat H_c\end{array}\right)\overline{\varphi}_\pm(x)e^{\pm
ik_x x}
$$
\begin{equation}\simeq
e^{\pm ik_x x}\left( \begin{array}{cc} i\omega_n -\hat H^{a}_\pm &
\Delta(x)\\ \Delta^*(x)& i\omega_n +\hat H^{a}_{\pm c}\end{array}\right)
\overline{\varphi}_\pm(x),
\end{equation}
where we defined $k_x=\sqrt{k_F^2-k_\parallel^2}$ and
\begin{equation}
\hat H^{a}_\pm=\mp
iv_x\partial_x-\frac{e}{c}\bbox{A_\parallel}(x)\bbox{v_\parallel} +
\frac{e^2}{2m c^2}\bbox{A_\parallel}^2(x) +\tilde V(x).
\end{equation}
Here  $v_x=k_x/m$, $\tilde V(x)$ represents a slowly varying part of the
potential which {\it does not} include fast variations which may occur at
metallic interfaces. The latter will be accounted for by the boundary
conditions to be formulated below. But first let us briefly describe
the general structure of the Green functions obeying eq. (\ref{start}).

\subsection{Construction of the Green functions}
Consider the equation
\begin{equation}\left(
\begin{array}{cc} i\omega_n -\hat H^{a}_{\pm} & \Delta(x)\\ \Delta^*(x)&
i\omega_n +\hat H^{a}_{\pm c} \end{array}\right) \overline{\varphi}_\pm=0.
\label{appr}
\end{equation}
There exist two linearly
independent solutions $\overline{\varphi}_+$ of eq. (\ref{appr}). One such
solution (denoted below by $\overline{\varphi}_{+1}$) does not diverge
at $x\rightarrow +\infty$, the other solution $\overline{\varphi}_{+2}$
is well-behaved at $x\rightarrow -\infty$.
Similarly, two linearly independent solutions $\overline{\varphi}_{-1,2}$
do not diverge respectively at $x\rightarrow -\infty$ and
$x\rightarrow +\infty$.

A particular solution of the
Gor'kov equations (\ref{start}) can now be sought in the following form

\begin{eqnarray}
\left( \begin{array}{c} G_{\omega_n} (x,x',\bbox{k_\parallel})\\F^+_{\omega_n}
(x,x',\bbox{k_\parallel})\end{array}\right)=\overline{\varphi}_{+1}(x) g_1(x')
e^{ik_x(x-x')}+ \nonumber\\\overline{\varphi}_{-2}(x) g_2(x')
e^{-ik_x(x-x')}\quad \mbox{if}\: x>x' \label{xbp}
\end{eqnarray}
and
\begin{eqnarray}
\left( \begin{array}{c} G_{\omega_n} (x,x',\bbox{k_\parallel})\\F^+_{\omega_n}
(x,x',\bbox{k_\parallel})\end{array}\right)=\overline{\varphi}_{-1}(x) f_1(x')
e^{-ik_x(x-x')}+ \nonumber\\\overline{\varphi}_{+2}(x) f_2(x')
e^{ik_x(x-x')}\quad \mbox{if}\: x<x'\label{xmp}.
\end{eqnarray}
These functions satisfy Gor'kov equations at $x\neq x'$. The functions
$f_{1,2}(x)$ and $g_{1,2}(x)$ are determined with the aid of the
continuity condition for the Green functions at $x=x'$ and the condition
resulting from the integration of $\delta(x-x')$ in eq.(\ref{start}).
As a result we arrive at the linear equations
\begin{eqnarray}
&& \overline{\varphi}_{+1}(x) g_1(x)+\overline{\varphi}_{-2}(x) g_2(x)=
\nonumber\\  &&\overline{\varphi}_{-1}(x) f_1(x)+\overline{\varphi}_{+2}(x)
f_2(x), \label{trr}\\ &&\frac{iv_x}{2}\Big[\overline{\varphi}_{+1}(x) g_1(x)-
\overline{\varphi}_{-2}(x) g_2(x)+ \nonumber\\&& \overline{\varphi}_{-1}(x)
f_1(x)-\overline{\varphi}_{+2}(x) f_2(x)\Big]=\left( \begin{array}{c}
1\\0\end{array}\right),\nonumber
\end{eqnarray}
which can be trivially resolved.

For a homogeneous superconductor in the absence of the
magnetic field this procedure allows to immediately recover
the well known result
\begin{eqnarray}
& G_{\omega_n}(x,x')=-\frac{i}{v_x(1+\gamma^2)}\left( e^{ik_S|x-x'|}-
\gamma^2e^{-ik_S^*|x-x'|}\right),&\nonumber\\
&F^+_{\omega_n}(x,x')=\frac{\gamma e^{-i\chi}}{ v_x(1+\gamma^2)}\left(
e^{ik_S|x-x'|}+e^{-ik_S^*|x-x'|}\right),&\nonumber
\end{eqnarray}
where $\chi$ is the phase of the pairing potential,
$k_S=k_x+i\Omega_n/v_x$,
$\Omega_n=\sqrt{|\Delta|^2+\omega_n^2}$ and $
\gamma=\frac{|\Delta|}{\omega_n+\Omega_n}$.
Here for convenience we set  $\omega_n>0$.

In a non-homogeneous situation a general solution of the
Gor'kov equations takes the form
\begin{eqnarray}
&\left( \begin{array}{c} G_{\omega_n} (x,x')\\F^+_{\omega_n} (x,x')
\end{array}\right)= \left( \begin{array}{c}
G_{\omega_n} (x,x')\\F^+_{\omega_n} (x,x')\end{array}\right)_{part}+&\nonumber
\\ &[ l_1(x')\overline{\varphi}_{+1}(x) +l_2(x')\overline{\varphi}_{+2}(x)]
e^{ik_x x}+& \label{general} \\ &[l_3(x')\overline{\varphi}_{-1}(x)+
l_4(x')\overline{\varphi}_{-2}(x)]e^{-ik_x x}.&\nonumber
\end{eqnarray}
For systems which consist of several metallic layers the
particular solution is obtained with the aid of the procedure outlined above
provided both coordinates $x$ and $x'$ belong to the same layer. Should
$x$ and $x'$ belong to different layers, the particular solution is zero
because in that case the $\delta$-function in eq. (\ref{start}) fails.
The functions $l_{1,2,3,4}(x')$ in each layer should be derived from the proper
boundary conditions which we will now specify.

\subsection{Boundary conditions}

In what follows we shall assume interfaces to be non-magnetic. In this case
matching of the wave functions on the left and on the right side of a
potential barrier, respectively $A_1 \exp(ik_{1x}x)+ B_1 \exp(-ik_{1x}x)$ and
$A_2 \exp(ik_{2x}x)+ B_2 \exp(-ik_{2x}x)$, is performed in a standard way (see
e.g \onlinecite{LL}):
\begin{eqnarray}
&& A_2=\alpha A_1+\beta B_1,\: B_2=\beta^*A_1 +\alpha^* B_1,\nonumber
\\
&& |\alpha|^2-|\beta|^2=\frac{k_{1x}}{k_{2x}}. \label{scatt}
\end{eqnarray}
The reflection and transmission coefficients are given by
\begin{equation}
R=\left|\frac{\beta}{\alpha}\right|^2,\quad
D=1-R=\frac{k_{1x}}{k_{2x}|\alpha|^2}. \label{scatt2}
\end{equation}
For the sake of simplicity below we shall set $k_{1x}=k_{2x}$.
Since typical energies of interest, such as $\Delta$ and typical Matsubara
frequencies, are all much smaller than the magnitude of the interface
potentials, the relationships (\ref{scatt}) can be directly applied to
the two-element columns in eq. (\ref{general}). In this way we
uniquely determine the Green functions of our problem.

For an illustration let us consider a metallic layer with the left and right
boundaries located respectively at $x=d_1$ and $x=d_2$. We will also choose
the argument $x'$ inside this layer. As it was already explained, the
particular solution of the Gor'kov equations for $x<d_1$ or $x>d_2$ is
$G(x,x')=F^+(x,x')=0$, while it has the form (\ref{xbp}), (\ref{xmp}) if the
coordinate $x$ belongs to this layer. Thus at the left boundary we get
\begin{eqnarray}
&&\overline{\varphi}_{+2}(d_1) f_2(x') e^{-ik_x x'}+
l_1(x')\overline{\varphi}_{+1}(d_1)
+l_2(x')\overline{\varphi}_{+2}(d_1)=\nonumber\\
&&\alpha_1\left[l_1^L(x')\overline{\varphi}_{+1}^L(d_1)
+l_2^L(x')\overline{\varphi}_{+2}^L(d_1)\right]+\nonumber\\ &&\beta_1\left[
l_3^L (x')\overline{\varphi}_{-1}^L(d_1)+ l_4^L (x')\overline{\varphi}_{-2}^L
(d_1)\right]\label{leftb1}
\end{eqnarray}
and
\begin{eqnarray}
&&\overline{\varphi}_{-1}(d_1) f_1(x') e^{ik_x x'}+
l_3(x')\overline{\varphi}_{-1}(d_1)+
l_4(x')\overline{\varphi}_{-2}(d_1)=\nonumber\\ &&\beta_1^*
\left[l_1^L(x')\overline{\varphi}_{+1}^L(d_1)
+l_2^L(x')\overline{\varphi}_{+2}^L(d_1)\right] +\nonumber\\ &&\alpha_1^*
\left[ l_3^L (x')\overline{\varphi}_{-1}^L(d_1)+ l_4^L
(x')\overline{\varphi}_{-2}^L (d_1)\right].\label{leftb2}
\end{eqnarray}
The superscript $L$ labels the solutions in the layer
located at $x<d_1$. The above boundary conditions provide
four linear equations for the functions $l(x')$ with the source
term $f_{1,2}(x')\exp(\pm ik_x x')$. Similarly, with the aid of
eq. (\ref{xbp}) four boundary conditions at the right boundary $x=d_2$
can be established. Analogous procedure should be applied to other
interfaces.

\section{Josephson current in $\bbox{SINI'S}$ junctions}

We shall consider $SNS$ junctions composed of clean superconducting
($S$) and normal ($N$) metals. We will assume that a thin insulating
layer ($I$) can be present at both $SN$ interfaces which, therefore, will
be characterized by arbitrary transparencies ranging from zero to one.
Specular reflection at both interfaces will be assumed. We also assume
that between interfaces electrons propagate ballistically and no
electron-electron or electron-phonon interactions are present in
the normal metal. For simplicity we will restrict our attention to the
case of identical superconducting electrodes with singlet
isotropic pairing. Furthermore, we shall neglect possible suppression
of the superconducting order parameter $\Delta$ in the electrodes
close to the $SN$ interface. This is a standard approximation which
is well justified in a large number of cases. The phase of the order
parameter is set to be $-\chi /2$ in the left electrode
and $\chi /2$ in the right one. The thickness of the normal layer
is denoted by $d$.

In order to evaluate the dc Josephson current across this structure
we shall follow the quasiclassical approach described
in the previous section. Technical details of our calculation are
presented in Appendix A. As a result we arrive at the expression
for the two point Green function in the normal layer. After that the
current density can be calculated from the standard formula
\begin{equation}
 J=\frac{ie}{m} T\sum_{\omega_n}\int
\frac{d^2k_\parallel}{(2\pi)^2}\left(\nabla_{x'}- \nabla_x\right)_{x'\to
x}G_{\omega_n} (x,x',\bbox{k_\parallel}). \label{ccuu}
\end{equation}
On can also rewrite the current in the form
 $J=J_+(\chi)-J_+^*(-\chi)$
where $J_+$ is defined by eq. (\ref{ccuu}) with positive
Matsubara frequencies $\omega_n >0$. Using (\ref{str}) and
omitting terms oscillating at atomic distances we obtain
\begin{equation}
J_+=2ie T\sum_{\omega_n>0} \int_{|k_\parallel|<k_F}
\frac{d^2k_\parallel}{(2\pi)^2}(V_1-U_2),
\end{equation}
or explicitly
\begin{equation}
J=4e T \sin\chi\sum_{\omega_n>0}\int_0^{k_F}
\frac{k_xdk_x}{2\pi} \frac{\sin\chi}{\cos\chi +W},
\label{J}
\end{equation}
where we defined
\begin{eqnarray}
W=\frac{4\sqrt{R_1R_2}}{D_1D_2}\frac{\Omega_n^2}{\Delta^2}\cos(2k_x d+\phi)
\nonumber
\\+\frac{\Omega_n^2
(1+R_1)(1+R_2)+ \omega_n^2D_1D_2}{D_1D_2\Delta^2}
\cosh\frac{2\omega_nd}{v_x}\label{W}\\
+\frac{2(1-R_1R_2)}{D_1D_2}
\frac{\Omega_n\omega_n}{\Delta^2}\sinh\frac{2\omega_nd}{v_x}.
\nonumber
\end{eqnarray}
Here $2k_xd+\phi$ is the phase of the product
$\alpha_2^*\beta_2\alpha_1^*\beta_1^*$. Eqs. (\ref{J}), (\ref{W})
provide a general expression for the dc Josephson current in $SINI'S$
structures valid for arbitrary transmissions $D_1$ and $D_2$ ranging
from zero to one. This expression is the central result of this
section.

We also note that the integral over $k_x$ in eq. (\ref{J}) can be
rewritten as a sum over independent conducting channels
\begin{equation}
\frac{{\cal A}}{2\pi}\int_0^{k_F}k_x dk_x(...) \to\sum_m^N (...),
\label{Dm}
\end{equation}
where ${\cal A}$ is the junction cross section. In this case
$D_{1,2}$ and $R_{1,2}$ may also depend on $m$. This would
correspond to different transmissions for different conducting channels.

Finally let us point out that in the limit of symmetrical
low transparent barriers $D_{1}=D_{2}\ll 1$
the problem was recently studied by Brinkman and Golubov\cite{brink}.
In the corresponding limit their result (eq. (8) of Ref.
\onlinecite{brink}) is similar -- although not fully equivalent --
to our eqs. (\ref{J}), (\ref{W}).

\subsection{Junctions with few conducting channels}

Let us first  analyze the above result for the case of one
conducting channel $N=1$. We observe that the first term in
eq. (\ref{W}) contains $\cos(2k_x d+\phi)$ which oscillates
at distances of the order of the Fermi wavelength. Provided at least
one of the barriers is highly transparent and/or
(for sufficiently long junctions $d \gtrsim \xi_0$) the temperature is high
$T \gg v_F/d$ this oscillating term is unimportant and can be
neglected. However, at lower transmissions of both barriers
and for relatively short
junctions $d \lesssim v_F/T$ this term turns out to be of the same order
as the other contributions to $W$ (\ref{W}). In this case
the supercurrent is sensitive to the exact positions of the discrete energy
levels inside the junction which can in turn vary considerably
if $d$ changes at the atomic scales $\sim 1/k_F$. Hence, one can
expect sufficiently strong sample-to-sample fluctuations of the
Josephson current even for junctions with nearly identical
parameters.

Let us first consider the limit of relatively short $SINI'S$
junctions in which case we obtain
\begin{equation}
I=\frac{e\Delta}{2}\frac{{\cal T}\sin \chi}{{\cal D}}
\tanh \left[\frac{{\cal D}\Delta }{2T}\right],
\label{I}
\end{equation}
where we defined
\begin{equation}
{\cal D}=\sqrt{1-{\cal T}\sin^2(\chi /2)}
\label{D1}
\end{equation}
and an effective normal transmission of the junction
\begin{equation}
{\cal T}=\frac{D_1D_2}{1+R_1R_2+2\sqrt{R_1R_2}\cos(2k_x d+\phi )}.
\label{T}
\end{equation}
Eq. (\ref{I}) has exactly the same functional form as the
result derived by Haberkorn {\it et al.} \cite{Hab}
for $SIS$ junctions with an arbitrary transmission
of the insulating barrier. This result is recovered from our eqs.
(\ref{I}), (\ref{T}) if we assume
e.g. $D_1 \ll D_2$ in which case the total transmission (\ref{T})
reduces to ${\cal T} \simeq D_1$.

As we have already discussed the total transmission ${\cal T}$ and,
hence, the Josephson current fluctuate depending on the exact position
of the bound states inside the junction. The resonant transmission is
achieved for $2k_xd +\phi=\pm\pi$, in which case we get
\begin{equation}
{\cal T}_{\rm res}=\frac{D_1D_2}{(1-\sqrt{R_1R_2})^2}.
\label{Tres}
\end{equation}
This equation demonstrates that for symmetric junctions $D_1=D_2$ at
resonance the Josephson current does not depend on the barrier
transmission at all. In this case ${\cal T}_{\rm res}=1$ and our
result (\ref{I}) coincides with the formula derived by Kulik and
Omel'yanchuk\cite{KO} for ballistic constrictions. In the limit
of low transmissions $D_{1,2} \ll 1$ we recover the standard
Breit-Wigner formula ${\cal T}_{\rm res}=4D_1D_2/(D_1+D_2)^2$
and reproduce the result obtained by Glazman and Matveev\cite{glazman}
for the problem of resonant tunneling through a single Anderson
impurity between two superconductors.

Note that our results (\ref{I}-\ref{T}) also support the conclusion
reached by Beenakker\cite{Ben} that the Josephson current across
sufficiently short junctions has a universal form and depends only on the
total scattering matrix of the weak link which can be evaluated in the
normal state. Although this conclusion is certainly correct in the
limit $d \to 0$, its applicability range depends significantly on the
physical nature of the scattering region. From eqs. (\ref{J}),
(\ref{W}) we observe that the result (\ref{I}), (\ref{D1}) applies
at $d \ll \xi_0$ not very close to the resonance. On the other hand,
at resonance the above result is
valid only under a more stringent condition
$d \ll \xi_0D_{\rm max}$, where we define $D_{\rm max}=$max$(D_1,D_2)$.

Now let us briefly analyze the opposite limit of sufficiently long junctions
$d\gg \xi_0$. Here we will restrict ourselves to the most interesting case
$T=0$. From eqs. (\ref{J}), (\ref{W}) we obtain
\begin{eqnarray}
&& I=\frac{ev_x\sin\chi}{\pi d
z_1}\left[\frac{\arctan\sqrt{z_2/z_1}}{\sqrt{z_2/z_1}}\right],
\\&& z_1=\cos^2(\chi/2)+\frac{1}{D_1 D_2}\left( R_++2\sqrt{R_1 R_2}
\cos(2k_x d+\phi) \right), \nonumber
\\ && z_2=\sin^2(\chi/2)+\frac{1}{D_1 D_2}\left(
R_+-2\sqrt{R_1 R_2} \cos(2k_x d+\phi)\right),\nonumber
\end{eqnarray}
where $R_+=R_1+R_2$. For a fully transparent channel
$D_1=D_2=1$ the above expression
reduces to the well known Ishii-Kulik result\cite{Ishii,Kulik}
\begin{equation}
I=\frac{ev_x\chi}{\pi d},\quad -\pi<\chi<\pi,
\end{equation}
whereas if one transmission is small $D_1 \ll 1$ and $D_2 \approx 1$
we reproduce the result\cite{ZZh}
\begin{equation}
I=\frac{ev_xD_1\sin\chi}{2d}.
\label{Z}
\end{equation}
Provided the transmissions of both $NS$-interfaces are low $D_{1,2}\ll
1$ we obtain in the off-resonant region
\begin{equation}
I=\frac{ev_x}{4\pi d}D_1D_2\sin\chi\Upsilon[2k_xd+\phi],
\end{equation}
where $\Upsilon[x]$ is a $2\pi$-periodic function defined as
\begin{equation}
\Upsilon[x]=\frac{x}{\sin{x}},\quad  -\pi<x<\pi.
\end{equation}
In the vicinity of the resonance $||2k_xd +\phi|-\pi
 |\lesssim D_{\rm max}$ the above result does not hold
anymore. Exactly at resonance $2k_xd +\phi=\pm \pi$ we get
\begin{equation}
I=\frac{ev_x\sqrt{D_1 D_2} \sin\chi}{4d
\left\{\cos^2\frac{\chi}{2}+\frac{1}{4}\left(
\sqrt{\frac{D_1}{D_2}}-\sqrt{\frac{D_2}{D_1}}\right)^2\right\}^{1/2}}.
\end{equation}
For a symmetric junction $D_{1,2}=D$ this formula yields
\begin{equation}
I=\frac{ev_xD \sin(\chi/2)}{2d},\quad  -\pi<\chi<\pi,
\end{equation}
while in a strongly asymmetric case $D_1 \ll D_2$ we again arrive
at the expression (\ref{Z}). This implies that at resonance the
barrier with higher transmission $D_2$ becomes effectively transparent
even if $D_2 \ll 1$. We conclude that for $D_{1,2} \ll 1$ the maximum
Josephson current is proportional to the product of transmissions
$D_1D_2$ off resonance, whereas exactly at resonance it is
proportional to the lowest of two transmissions $D_1$ or $D_2$.

We observe that both for short and long $SINI'S$ junctions interference
effects may enhance the Josephson effect or partially suppress it
depending on the exact positions of the bound states inside the junction. We
also note that in order to evaluate the supercurrent across $SINI'S$ junctions
it is in general {\it not} sufficient to derive the transmission probability
for the corresponding $NINI'N$ structure. Although the normal
transmission of the above structure is given by eq. (\ref{T}) for {\it all}
values of $d$, the correct expression for the Josephson current can be
recovered by combining eq. (\ref{T}) with the results\cite{Hab,Ben}
in the limit of short junctions $d \ll D\xi_0$ only. In this case one can
neglect suppression of the anomalous Green functions inside the normal layer
and, hence, the information about the normal transmission turns out to be
sufficient. On the contrary, for longer junctions the decay of Cooper pair
amplitudes inside the $N$-layer cannot be anymore disregarded. In this case
the supercurrent will deviate from the form (\ref{I}) even though the normal
transmission of the junction (\ref{T}) will remain unchanged. This deviation
becomes particularly pronounced for long junctions, i.e. for $d \gg \xi_0$ out
of resonance and for $d \gg D\xi_0$ at resonance.

The above analysis can trivially be generalized to the case of an
arbitrary number of independent conducting channels inside the
junction $N >1$. In that case the supercurrent is simply given by
the sum of the contributions from all the channels. Although
all these contributions have the same form, they are in general
not equal because the phase factors $2k_xd+\phi$ change randomly
for different channels. Accordingly, mesoscopic fluctuations
of the supercurrent should become smaller with increasing number
of channels and eventually disappear in the limit of large $N$.
In the latter limit the Josephson current is obtained by
averaging over all values of the phase $2k_xd+\phi$. The corresponding
results are presented below.

\subsection{Many channel limit}

Averaging over the phase factors $2k_xd+\phi$ is effectively performed by
integrating over directions of the electron momentum in eq. (\ref{J}). Since
the term in the expression for $W$ (\ref{W}) which contains $\cos(2k_x
d+\phi)$ oscillates very rapidly with changing $k_x$, averaging can be
performed by first integrating the current (\ref{J}) over the phase
$2k_xd+\phi$ and then integrating the result over $k_x$. We obtain
\begin{equation}
J=\frac{2}{\pi}ek_F^2 T \sin\chi\sum_{\omega_n>0}\int_0^1\mu
d\mu \frac{t_1(\mu )t_2(\mu )}{{\cal Q}^{1/2}(\chi , \mu)}.
\label{JN}
\end{equation}
Here and below we define $\mu =k_x/k_F$,
$
t_{1,2}(\mu)=D_{1,2}(\mu)/(R_{1,2}(\mu)+1)
$,
$t_\pm =t_1\pm t_2$ and
\begin{eqnarray}
\label{Q}
{\cal Q}=\left[t_1t_2\cos\chi+
\big(1+(t_1t_2+1)\frac{\omega_n^2}{\Delta^2} \big) \cosh\frac{2\omega_nd}{\mu
v_F}\right.\\
\left.+t_+\frac{\omega_n\Omega_n}{\Delta^2}
\sinh\frac{2\omega_nd}{\mu v_F}\right]^2-
(1-t_1^2)(1-t_2^2)\frac{\Omega_n^4}{\Delta^4}.
&\nonumber
\end{eqnarray}
The above equations fully determine the Josephson current in $SINI'S$
junctions in the many channel limit and at arbitrary transmissions of
specularly reflecting $SN$-interfaces.

Let us make use of this result in order to perform a direct comparison between
our analysis and the approach based on the Eilenberger equations supplemented
by Zaitsev boundary conditions. The corresponding calculation within the
latter approach is performed in Appendix B. It is interesting to observe that
for $SINI'S$ junctions this calculation yields exactly the same result
(\ref{JN}), (\ref{Q}) as obtained within our calculation after averaging over
the scattering phase $2k_xd +\phi$.

\begin{figure}
\centerline{\epsfclipon\epsfxsize=1.0\hsize\epsffile{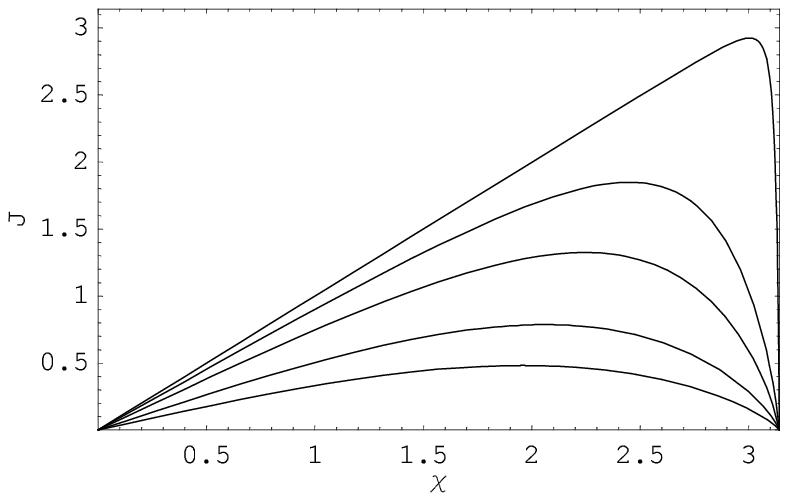}} {\small
Fig. 1. The Josephson current density (\ref{Jour}) normalized by
$ek_F^2v_F/6\pi^2 d$ is plotted as a function of the phase difference $\chi$.
Here we assumed that the boundary is described by an effective potential
$U_0\delta(x\pm d/2)$ in which case one has $t=v_x^2/(2U_0^2+v_x^2)$. The
dependence $J(\chi )$ was evaluated for $2U_0^2/v_F^2=10^{-4},
0.03,0.1,0.3,0.6$ (from top to bottom).}
\end{figure}

This observation allows to make an important conclusion concerning the
applicability of the quasiclassical analysis employing Zaitsev boundary
conditions for the Eilenberger propagators. The exact result for the Josephson
current in $SINI'S$ systems, eqs. (\ref{J}), (\ref{W}), cannot be recovered
within the latter approach because it essentially ignores interference effects
arising from electron scattering on two insulating barriers. At the same time,
in the limit of many conducting channels the scattering phase is effectively
averaged out. In this limit Zaitsev boundary conditions turn out to correctly
describe the supercurrent. It is also important to emphasize that the latter
conclusion applies for the systems with not more than two barriers. Below we
will analyze the supercurrent in $SNS$ structures with three insulating
barriers and will show that the approach based on Zaitsev boundary conditions
fails to provide correct results even in the limit of many conducting
channels.

But first let us present several limiting expressions for the sake of
completeness. We start from the limit of a sufficiently thick junction $d\gg
\xi_0$ and consider $T=0$. In this case we find
\begin{equation}
J=\frac{ek_F^2 v_F\sin\chi}{2 d\pi^2}\int_0^1  d\mu \mu^2
t_1t_2\sqrt{f_1f_2} F(\varphi, h),
\label{ell}
\end{equation}
where $F(\varphi,h)=\int_0^\varphi(1-h\sin^2\theta)^{-1/2}d\theta$ is the
incomplete elliptic integral, $\varphi =\arcsin (1/\sqrt{f_1})$,
$h=f_1+f_2-f_1f_2$ and
$$ f_1=\frac{1}{1-t_1t_2\cos^2\frac{\chi}{2}-
\frac{1}{2}\big[1-t_1t_2-\sqrt{(1-t_1^2)(1-t_2^2)}\big]},
$$
$$
f_2=\frac{1}{1-t_1t_2\sin^2\frac{\chi}{2}-
\frac{1}{2}\big[1-t_1t_2-\sqrt{(1-t_1^2)(1-t_2^2)}\big]}.
$$
For an $SINS$ junction ($t_2=1$) the above result yields (cf. Ref.
\onlinecite{svi})
$$
J=\frac{ek_F^2 v_F\sin\chi}{d\pi^2}\int_0^1 \frac{d\mu
\mu^2t_1}{\sqrt{1-t_1^2\cos^2\chi}}\arctan
\sqrt{\frac{1-t_1\cos\chi}{1+t_1\cos\chi}}.$$ The expression (\ref{ell}) also
simplifies in the case of a symmetric junction $t_1=t_2$
\begin{equation}
J=\frac{ek_F^2 v_F}{\pi^2
d}\int_0^1 \frac{\rho\mu^2 d\mu}{\sqrt{1+\rho^2}}
F\left(y,\frac{1}{1+\rho^2}\right),
\label{Jour}
\end{equation}
where
$$
y=\arccos\left[t\cos(\chi/2)\right],\;\;\;\;
\rho(\mu,\chi)=\frac{t^2\sin\chi}{2\sqrt{1-t^2}}.
$$
The current density $J$ (\ref{Jour}) is plotted in
Fig. 1 as a function of the Josephson phase $\chi$
for several values of the barrier transmission.
Note that in the case of small interface transparencies
the limit $T \to 0$ is effectively achieved at temperatures
much lower than $tv_F/d$.

Let us now proceed to the case of small transparencies of both interfaces
$t_{1,2}\ll 1$. In this limit the expression (\ref{Q}) takes the form
\begin{eqnarray}
{\cal Q}=\frac{\Omega_n^4}{\Delta^4} \left[ \sinh\frac{2\omega_n d}{\mu
v_F}+t_+\frac{\omega_n}{\Omega_n}\right]^2+\frac{\Omega_n^2}{\Delta^2}{\cal
P}(\mu, \chi), \label{qsm}
\end{eqnarray}
where
\begin{equation}
{\cal P}(\mu, \chi)= t_+^2(\mu)\cos^2(\chi/2)+
t_-^2(\mu)\sin^2(\chi/2).
\end{equation}
As we have already pointed out, the above result is not identical
to one presented in eq. (13) of Ref. \onlinecite{brink}
(see also \cite{Misha}). However, it
is easy to see that this difference does not affect the final expression
for the current in two important limits of short ($d\ll t\xi_0$)
and long ($d\gg t\xi_0$) junctions. Only in the intermediate case $d\sim
t\xi_0$ some deviations between our results and those of Ref.
\onlinecite{brink} are observed. This is demonstrated in Fig.2.

The case of short junctions $d\ll t\xi_0$ was already
studied in Ref. \onlinecite{brink}.
Therefore here we only present the asymptotic expression for the current at
$d\gg t\xi_0$
\begin{eqnarray}
J=\frac{ek_F^2v_F\sin\chi}{2\pi^2
d}\int_0^1 d\mu\mu^2 t_1t_2\ln(\epsilon_1/\epsilon_2),
\label{dras}
\end{eqnarray}
where $\epsilon_1=\mbox{min}\{\mu v_F/d,\Delta\}$, $\epsilon_2=\mu
v_F/(4d\sqrt{{\cal P}})$ for $T\ll tv_F/d$ and $\epsilon_2\simeq T$ for
$tv_F/d \ll T \ll \epsilon_1$.
The accuracy of the above formula is in general logarithmic, and it
becomes next to logarithmic in the limits $d\ll \xi_0$ or
$d\gg \xi_0$.

\begin{figure}
\centerline{\epsfclipon\epsfxsize100mm\epsffile{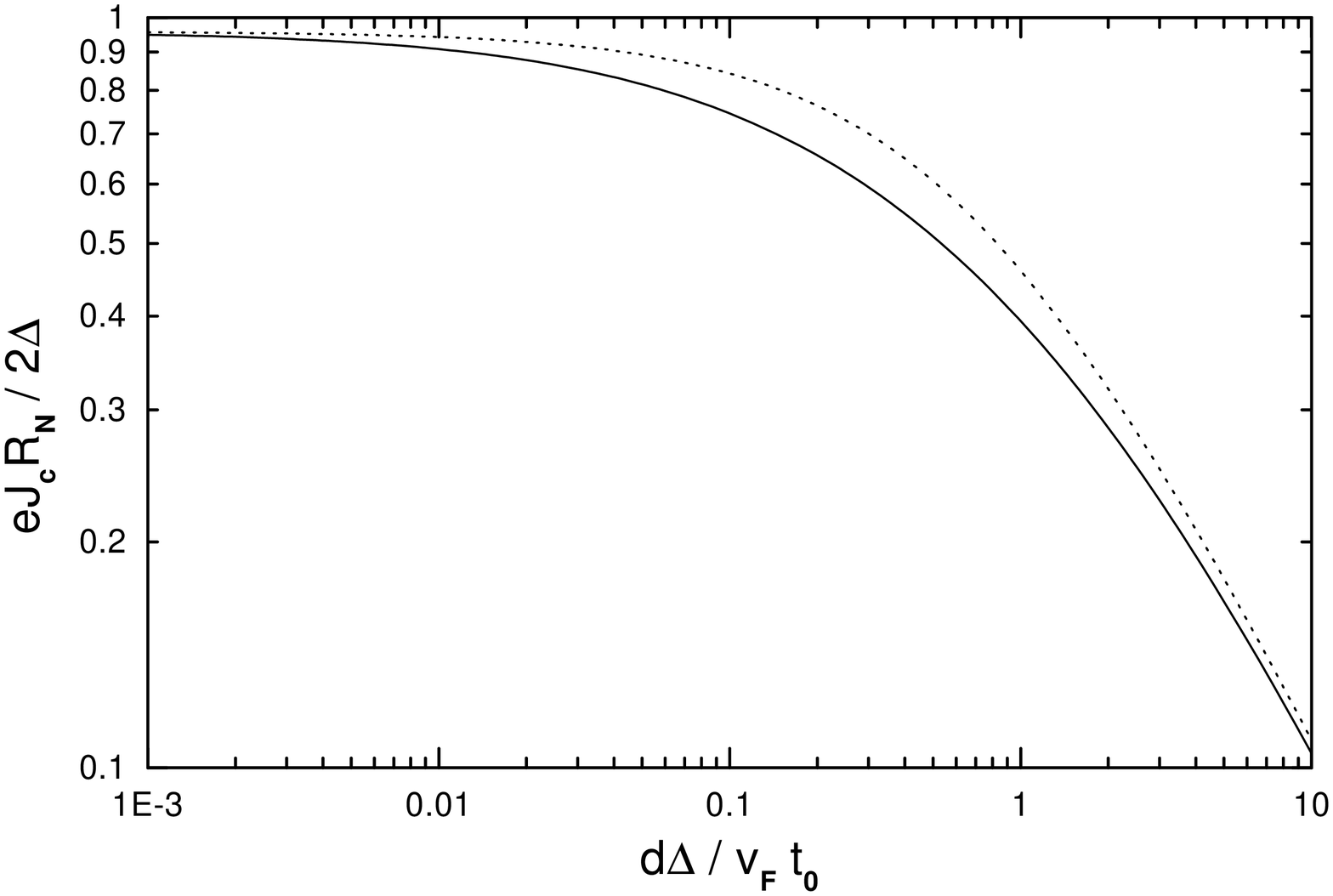}} {\small Fig. 2.
The maximum Josephson current $J_c$ of a symmetric $SINIS$-junction with low
transparency. The angular dependence of transparency is taken to be
$t(\mu)=t_0\mu^2$. The normal-state resistance of the structure is denoted by
$R_N$. The solid line represents eqs. (\ref{JN}), (\ref{qsm}) of our paper and
the dashed line represents eq. (13) of Ref. [18]. }
\end{figure}

\section{Josephson current in $\bbox{SINI'NI''S}$ junctions}

Let us now consider $SNS$ structure with three insulating barriers. As
before, two of them are located at $SN$ interfaces, and the third
barrier is inside the $N$-layer at a distance $d_1$ and $d_2$
respectively from the left and right $SN$ interfaces.
The transmission and reflection coefficients of this intermediate
barrier are denoted as $D_0$ and $R_0=1-D_0$, whereas the left and
the right barriers are characterized respectively by $D_1=1-R_1$ and
$D_2=1-R_2$.

The supercurrent is evaluated along the same lines as it was done in
Section 2 for the case of two barriers. A straightforward (although
somewhat lengthy and cumbersome) procedure yields the final result for the
Josephson current which is again expressed by eq. (\ref{J}),
where the function $W$ now takes the form
\begin{equation}
W=W_++W_-+W_{12} \label{wsum}
\end{equation}
These three contributions to the $W$-function depend respectively
on the sum of thicknesses $d_1$ and $d_2$, their difference and on these
two values separately. We find
\end{multicols}
\begin{eqnarray}
& W_{+}=\frac{4\sqrt{R_1R_2}}{D_0D_1D_2}\frac{\Omega_n^2}{\Delta^2}
\cos[2k_x (d_1+d_2)+\phi_1+\phi_2]+
\frac{2(1-R_1R_2)}{D_0D_1D_2}\frac{\Omega_n\omega_n}{\Delta^2}
\sinh\frac{2\omega_n(d_1+d_2)}{v_x} &\nonumber
\\ & +\frac{\Omega_n^2
(1+R_1)(1+R_2)+ \omega_n^2D_1D_2}{D_0D_1D_2\Delta^2}
\cosh \frac{2\omega_n(d_1+d_2)}{v_x};&
\\ & W_{-}=\frac{4R_0\sqrt{R_1R_2}}{D_0D_1D_2}\frac{\Omega_n^2}{\Delta^2}
\cos[2k_x (d_1-d_2)+\phi_1-\phi_2]+
\frac{2R_0(D_1-D_2)}{D_0D_1D_2}\frac{\Omega_n\omega_n}{\Delta^2}
\sinh\frac{2\omega_n(d_1-d_2)}{v_x}
&\nonumber
\\ & +\frac{R_0}{D_0}\left[1+\frac{2\Omega_n^2}{\Delta^2}
\frac{R_1+R_2}{D_1D_2} \right]\cosh \frac{2\omega_n(d_1-d_2)}{v_x};
\\ & W_{12}=\frac{4\sqrt{R_0R_1}}{D_0D_1}\cos[2k_x d_1+\phi_1]\left[
\frac{\Omega_n\omega_n}{\Delta^2}\sinh\frac{2\omega_n
d_2}{v_x}+\frac{1+R_2}{D_2}\frac{\Omega_n^2}{\Delta^2}
\cosh\frac{2\omega_n d_2}{v_x} \right]&\nonumber
\\  & +\frac{4\sqrt{R_0R_2}}{D_0D_2}\cos[2k_x d_2+\phi_2]\left[
\frac{\Omega_n\omega_n}{\Delta^2}\sinh\frac{2\omega_n
d_1}{v_x}+\frac{1+R_1}{D_1}\frac{\Omega_n^2}{\Delta^2}
\cosh\frac{2\omega_n d_1}{v_x} \right].&
\end{eqnarray}
\begin{multicols}{2}
Here we introduced two phases
\begin{eqnarray}
2k_xd_{1}+\phi_{1}=\mbox{arg}\,\alpha_0^*\beta_0\alpha_{1}^*\beta_{1}^*,\\
2k_xd_{2}+\phi_{2}=\mbox{arg}\,\alpha_0^*\beta_0^*\alpha_{2}^*\beta_{2},
\nonumber
\end{eqnarray}
related to the corresponding elements of the scattering matrices for all three
barriers. In contrast to the case of two barriers these phases cannot be
simultaneously removed by shifting $k_x$. The expression for the Josephson
current in $SINI'S$ junctions derived in the previous section can easily be
recovered if we set $D_0=1-R_0=1$. By setting $D_{1,2}=1-R_{1,2}=0$ in the
above equations we arrive at the result for the supercurrent in $SNINS$
systems derived in Ref. \onlinecite{ZZh}.

\subsection{One channel limit}
Let us first analyze the above general result in the limit of one conducting
channel. In the limit of short junctions $d\ll \xi_0D_{\rm max}$ we again
reproduce the result (\ref{I}) where the total effective transmission of the
normal structure with three barriers takes the form
\begin{equation}
{\cal T}=\frac{2t_1t_0t_2}{1+t_1t_0t_2+{\cal
C}(\varphi_{1,2},t_{0,1,2})},
\end{equation}
where
$$
{\cal C}=\cos\varphi_1\sqrt{(1-t_0^2)(1-t_1^2)}
+\cos\varphi_2\sqrt{(1-t_0^2)(1-t_2^2)}
$$
\begin{equation}
+(\cos\varphi_1\cos\varphi_2-t_0\sin\varphi_1\sin\varphi_2)
\sqrt{(1-t_1^2)(1-t_2^2)}.
\end{equation}
Here we define $t_{0,1,2}=D_{0,1,2}/(1+R_{0,1,2})$ and
$\varphi_{1,2}=2k_xd_{1,2}+\phi_{1,2}$. For later purposes let us also perform
averaging of this transmission over the phases $\varphi_1$ and $\varphi_2$. We
obtain
\begin{equation}
\langle {\cal T}\rangle=\frac{2 t_1t_0 t_2}{\sqrt{2  t_1t_0 t_2
+t_1^2t_0^2+t_1^2t_2^2+t_0^2t_2^2-t_1^2t_0^2t_2^2}}.
\label{3t}
\end{equation}
In particular, in the case of similar barriers with small
transparencies $D_{0,1,2}\approx D \ll 1$ the average normal
transmission of our structure is  $\langle {\cal T}\rangle \sim
D^{3/2}$. Suppression of the average transmission below the value
$\sim D$ is a result of destructive interference and indicates
the tendency of the system towards localization. Eq. (\ref{3t})
follows from an explicit integration, but it can also be
understood in simple terms.  Consider the square $0<\varphi_1<2\pi,
0<\varphi_2<2\pi$. The main contribution to the average transmission
comes from the resonant region ${\cal T} \sim 1$. In the symmetric
case $t_{0,1,2}=t\ll 1$ this resonance occurs approximately along the
lines $\left[
\sqrt{(1+\cos\varphi_1)(1+\cos\varphi_2)}-t\right]^2\sim t^3$ in two quadrants
$\varphi_1,\varphi_2<\pi$ and $\varphi_1,\varphi_2>\pi$. In other
words, the resonant region is represented by two
hyperbola-like curves with characteristic widths $\sim
D^{3/2}$. This dependence of the average transmission is recovered from
the exact result (\ref{3t}).

Let us now proceed to the limit of a long junction $d_{1,2}\gg \xi_0$
and $T=0$. In the off-resonant region we find
\begin{equation}
I=\frac{ev_x D_1D_0D_2 \sin\chi}{8\pi d_1} {\cal B}(\varphi_{1,2}, d_2/d_1),
\label{I55}
\end{equation}
where
\begin{equation}
{\cal B}=\int_0^\infty \frac{dx}{\left[\cosh
x+\cos\varphi_1\right]\left[ \cosh (d_2 x/d_1)+\cos\varphi_2\right]}.
\label{B55}
\end{equation}
Evaluating this integral for $d_1=d_2$ we get
\begin{equation}
J=\frac{ev_x D_1D_0D_2 \sin\chi}{8\pi d_1}
\frac{\Upsilon[\varphi_1]-\Upsilon[\varphi_2]}{\cos\varphi_2-\cos\varphi_1}.
\end{equation}
This expression diverges at resonance (i.e. at $\varphi_1\simeq \pi$ or
$\varphi_2 \simeq \pi$) where it becomes inapplicable. In the resonant region
$\varphi_2 \simeq \pi$ we obtain
\begin{equation}
I=\frac{ev_x \sqrt{D_1D_0 D_2}\sin\chi}{4d
\sqrt{2(1+\cos\varphi_1)({\cal T}^{-1}-\sin^2(\chi/2))}}.
\end{equation}

\subsection{Many channel junctions}
As it was already discussed, in the many channel limit it is
appropriate to average the current over the scattering phases.
If the widths $d_1$ and $d_2$ fluctuate independently
on the atomic scale, averaging
over $\varphi_1$ and $\varphi_2$ can also be performed independently.
If $d_1$ and $d_2$ do not change on the atomic scale but
are incommensurate, independent averaging over the two phases can be performed
as well. The situation is different only for strictly
commensurate $d_1$ and $d_2$ in which case no independent averaging
can be fulfilled.

Let us first briefly discuss the latter situation of commensurate
$N$-layers. For simplicity we assume $d_1=d_2$, consider a symmetric situation
$D_1=D_2=D \ll 1$ and set the transparency of the intermediate interface to be
$D_0\gg D^2$. We will only present the result for the case of short junctions
$d\ll \xi_0D_{\rm max}$. We observe that the denominator in eq.
(\ref{J}), (\ref{wsum}) has a resonant structure as a function of
$\varphi_1+\varphi_2$. Integrating near the resonances we obtain
\begin{equation}
J= \frac{1}{2\pi}ek_F^2 T \sin\chi \int_0^1 d\mu \mu D(\mu)
\sum_{\omega_n>0}\frac{\Delta^2}{\Omega_n^2 \sqrt{{\cal R}(\chi,
\phi)}},
\end{equation}
where
\begin{equation}
{\cal R}=
(1+|\beta_0|^2\sin^2\phi)^2\left(1-\frac{\Delta^2\sin^2(\chi/2)}{
\Omega_n^2(1+|\beta_0|^2\sin^2\phi )}\right) \label{newsh}
\end{equation}
and $|\beta_0|^2(\mu)=[1-D_0(\mu)]/D_0(\mu)$. An
interesting feature of the expression (\ref{newsh}) is a dependence
of the critical Josephson current on the scattering phase
$\phi=(\phi_1-\phi_2)/2$. For instance, provided
$|\beta_0|$ is large (the transparency of the intermediate layer is small),
the critical current can vary from $\sim e k_F^2 |\Delta| D$ to $\sim e k_F^2
|\Delta| D D_0$ with $\phi$ varying from 0 to $\pi/2$. One should also
bear in mind that $\phi$ may depend on $\mu$. However, since
the main contribution to the supercurrent comes predominantly
from the electrons with the momenta perpendicular to the
interfaces, we can estimate the current with $\phi$ corresponding to the
forward direction. If $D_0\ll D^2$ and  $\sin^2\phi\gg D_0$ the current is
given by eq. (\ref{newsh}). For $\phi=0$ (or
$\phi=\pi$) a different expression follows
\begin{equation}
J=\frac{1}{2\pi} e k_F^2\sin\chi \Delta\tanh\frac{\Delta}{2T} \int_0^1 d\mu
\mu \frac{D_0(\mu)}{D(\mu)}.
\end{equation}

Now let us consider a more realistic
situation of incommensurate $d_1$ and $d_2$ which allows for
independent
averaging over the scattering phases $\varphi_1$ and $\varphi_2$.
Technically this procedure amounts to evaluating the integral
of the expression $1/[t+\cos x\cos(\lambda x)]$ from $x=0$ to some
large value $x=L$. At $\lambda=1$ the result of this
integration is $L/\sqrt{t(1+t)}$.
However, if $\lambda$ is irrational, the integral approaches the value
$2LK(1/t^2)/\pi t$, where $K(h)=F(\pi/2,h)$ is the complete elliptic
integral. This simple example illustrates our averaging procedure
over two independent phases $x$ and $\lambda x$.

Let us assume that the transparencies of all three interfaces are
small as compared to one. After
averaging over $\varphi_1$ one arrives the expression which has a resonant
dependence on $\varphi_2$ near $\varphi_2=\pi$. Expanding in powers
of near this resonance
with $\delta\varphi_2=\varphi_2-\pi$ and keeping the terms
proportional to $\delta\varphi_2^2$ and $\delta\varphi_2^4$ we find
\begin{eqnarray}&
\left\langle\frac{1}{\cos\chi+W_++W_-+W_{12}}\right\rangle_{\varphi_1}=&
\nonumber\\ & \frac{\Delta^2}{2\Omega_n^2} \left\{\frac{1}{D_1^{2}}+
\frac{2\delta \varphi_2^2}{D_0D_1D_2}
\left[1-\frac{2\Delta^2\sin^2(\chi/2)}{\Omega_n^2}\right]
+\frac{\delta\varphi_2^4}{D_0^2D_2^2} \right\}^{-1/2}. &\nonumber
\end{eqnarray}
Then evaluating the integral over $\delta\varphi_2$
we derive the final expression for the current
\begin{equation}
J=\frac{ek_F^2}{\pi^2}D_{\rm eff} \sin\chi T \sum_{\omega_n>0}\frac{\Delta
^2}{\Omega_n^2}K\left[ \frac{\Delta^2\sin^2(\chi/2)}{
\Omega_n^2}\right],
\label{kll}
\end{equation}
where we define the effective transmission
\begin{equation}
D_{\rm eff}=\int_0^1 \mu d\mu \sqrt{D_0D_1D_2}.
\label{Deff}
\end{equation}
Hence, for similar barriers we obtain the dependence $J \propto
D^{3/2}$ rather than $J \propto D$ (as it would be the case for independent
barriers). The latter dependence would follow from the calculation
based on Zaitsev boundary conditions for the Eilenberger propagators.
We observe, therefore, that quantum interference effects {\it
decrease} the Josephson current in systems with three insulating
barriers. This is essentially quantum effect which cannot be recovered
from Zaitsev boundary conditions even in the multichannel limit.
This effect has exactly the same origin as a quantum suppression of
the average normal transmission $\langle {\cal T} \rangle$ due to
localization effects. Further limiting expressions for short junctions
can be directly recovered from eq. (\ref{3t}).

We also note that the current-phase relation (\ref{kll}) deviates from
a pure sinusoidal dependence even though all three transmissions are small
$D_{0,1,2} \ll 1$.  At $T=0$ the critical
Josephson current is reached at $\chi \simeq 1.7$ which is slightly higher
than $\pi/2$. Although this deviation is quantitatively not very
significant, it is nevertheless important as yet one more indication
of quantum interference of electrons inside the junction.

Finally, let us turn to the limit of long junctions $d_{1,2}\gg\xi_0$. We
again restrict ourselves to the case of low transparent interfaces. At high
temperatures $T\gg v_F/2\pi d_{1,2}$  from eq. (\ref{J}),(\ref{wsum}) we get
\begin{equation}
J=\frac{eT  k_F^2}{\pi}\frac{\Delta^2\sin\chi}{\Delta^2+\pi^2 T^2}
\int_0^1 d\mu \mu D_0D_1D_2e^{-\frac{d}{\xi(T)\mu}},
\label{lds}
\end{equation}
where $d=d_1+d_2$ and $\xi(T)=v_F/(2\pi T)$. In this case the anomalous
 Green function strongly decays deep in the normal layer. Hence,
interference effects are not important and the interfaces can be
considered as independent from
each other. In the opposite limit $T\to 0$ (more precisely $T\ll Dv_F/d$),
however, interference effects become important, and the current becomes
proportional to $D^{5/2}$ rather than to $D^3$. Explicitly, at $T \to 0$ we
get
\begin{equation}
J=\frac{e k_F^2 v_F\sin\chi}{16\pi^2\sqrt{d_1d_2}}\int_0^1d\mu \mu^2
D_1D_2\sqrt{D_0}\ln D_0^{-1}.
\label{inc}
\end{equation}
This expression is valid with the logarithmic accuracy and
no distinction between $\ln D_0$, $\ln D_1$ or
$\ln D_2$ should be made. We see that, in contrast to short junctions,
in the limit of thick
normal layers interference effects {\it increase} the Josephson
current as compared to the case of independent barriers. The
result (\ref{inc}), as well as one of eqs. (\ref{kll}) (\ref{Deff}) cannot be
obtained from the Eilenberger approach
supplemented by Zaitsev boundary conditions.

\section{Discussion and conclusions}
Let us summarize our key results and observations.

In the present work we considered an interplay between the proximity
effect and quantum interference of electrons in hybrid structures
composed of normal metallic layers and superconductors. Quantum
interference effects occur between electrons scattered at different
metallic interfaces or other potential barriers and can strongly
influence the supercurrent across the system.

The standard quasiclassical approach which describes scattering at interfaces
by means of the nonlinear boundary conditions \cite{zait} for
energy-integrated Eilenberger propagators -- while very efficient in numerous
other situations -- is in general not suitable for the problem in question.
Because of this reason we made use of an alternative quasiclassical approach
which allows to investigate superconducting systems with more than one
potential barrier and fully accounts for the interference effects. Within this
approach scattering at boundaries is described with the aid of linear boundary
conditions for quasiclassical amplitude functions. Electron propagation
between boundaries is described by linear quasiclassical equations. Our
approach is technically not equivalent to one based on the Bogolyubov-de
Gennes equations. In particular, our method allows to explicitly construct
two-point Green functions of the system and bypass such intermediate steps as
finding an exact energy spectrum of the system with subsequent summation over
the energy eigenvalues inevitable within the Bogolyubov-De Gennes approach. On
the other hand, if needed, the full information about the energy bound states
can easily be recovered within our technique by finding the poles of the Green
functions in the Matsubara frequency plane.

Within our method we evaluated the dc Josephson current in $SNS$
junctions containing two and three insulating barriers with
arbitrary transmissions, respectively $SINI'S$ and
$SINI'NI''S$ junctions. For the system with two barriers and few
conducting channels we found
strong fluctuations of the Josephson critical current
depending on the exact position of the resonant level inside
the junction. For short junctions $d \ll \xi_0D$ at
resonance the Josephson current does not depend on the barrier
transmission $D$ and is given by the standard Kulik-Omel'yanchuk
formula \cite{KO} derived for ballistic weak links. In the limit
of long $SNS$ junctions $d \gg \xi_0$ resonant effects may also
lead to strong enhancement of the supercurrent, in this
case at $T \to 0$ and at resonance the Josephson current is proportional to $D$
and not to $D^2$ as it would be in the absence of interference
effects.

It is also interesting to observe that, while the above
results for few conducting channels cannot be obtained by means
of the approach employing Zaitsev boundary conditions, in the many
channel limit and for junctions with two barriers the latter
approach {\it does} allow to recover correct results. This is
because the contributions sensitive to the scattering phase
are effectively averaged out during summation over conducting
channels or, which is the same, during averaging of the current
over the directions of the Fermi velocity.

Quantum interference effects turn out to be even more
important in the proximity systems which contain three insulating
barriers. In this case the quasiclassical approach based on Zaitsev
boundary conditions fails even in the limit of many conducting
channels. In that limit the Josephson current is {\it decreased}
for short junctions ($J \propto D^{3/2}$) as compared to the
case of independent barriers ($J \propto D$). This effect is caused by
destructive interference of electrons reflected from different
barriers and indicates the tendency of the system towards
localization. In contrast, for long $SNS$ junctions with three
barriers an interplay between quantum interference and proximity
effect leads to enhancement of the Josephson current at $T \to 0$:
We obtained the dependence $J \propto D^{5/2}$ instead of $J \propto
D^3$ for independent barriers. We also discuss some further concrete
results which turn out to be quite sensitive to the details of the model.

Finally, we note that in a very recent publication\cite{shel2} Ozana and
Shelankov analyzed the applicability of the quasiclassical technique for the
case of superconducting sandwiches with several insulating barriers. For such
systems they also arrived at the conclusion that in the many channel limit the
standard quasiclassical scheme based on the Eilenberger equations {\it and}
Zaitsev boundary conditions effectively breaks down in the presence of more
than two reflecting interfaces. These authors argued that in such cases this
scheme disregards certain classes of interfering quasiclassical paths. This
conclusion\cite{shel2} is similar to one reached in the present paper for
$SNS$ structures. We would like to point out, however, that from our point of
view the failure of the above scheme is not so much due to the
quasiclassical approximation
and/or normalization conditions employed within the Eilenberger formalism.
The problem is rather in the boundary conditions\cite{zait} which
disregard interference effects which occur in the structures with
several interfaces/barriers with transmissions smaller than one.

We would like to thank J.C. Cuevas, D.S. Golubev, A.A. Golubov, M. Eschrig,
A. Shelankov, G. Sch\"on and U. Z\"ulicke for discussions and useful remarks.
The work is part of the {\bf CFN} (Center for
Functional Nanostructures) which is supported by the DFG (German Science
Foundation). We also acknowledge partial support of RFBR under Grant No.
00-02-16202.

\end{multicols}

\appendix

\section{}
Let us consider an $SINI'S$ system and assume that the normal metal layer is
located at $-d/2<x<d/2$. It is convenient to choose the coordinate $x'$ within
the normal layer, $-d/2<x'<d/2$. Then a general solution of eq. (\ref{start})
(decaying at $x \to -\infty$) in the left superconductor reads
\begin{equation}
\left( \begin{array}{c} G_{\omega_n} (x,x')\\F^+_{\omega_n}
(x,x')\end{array}\right)=  \left( \begin{array}{c} 1\\-i
e^{i\chi/2}\gamma^{-1}
\end{array}\right)e^{\kappa x/v_x}e^{ik_x x} f(x')+
\left( \begin{array}{c} 1\\i e^{i\chi/2}\gamma
\end{array}\right)e^{\kappa x}e^{-ik_x x} g(x'). \label{ls}
\end{equation}
Here $\kappa=\Omega_n/v_x$. The solution in the right superconductor can found
analogously. We get
\begin{equation}
\left( \begin{array}{c} G_{\omega_n} (x,x')\\F^+_{\omega_n}
(x,x')\end{array}\right)=  \left( \begin{array}{c} 1\\i e^{-i\chi/2}\gamma
\end{array}\right)e^{-\kappa x}e^{ik_x x} n(x')+
\left( \begin{array}{c} 1\\-i e^{-i\chi/2}\gamma^{-1}
\end{array}\right)e^{-\kappa x}e^{-ik_x x} r(x').
\label{rs}
\end{equation}
The above solutions contain four unknown functions $f(x')$, $g(x')$, $n(x')$
and $r(x')$. These functions should be found by matching (\ref{ls}),
(\ref{rs}) with the solution of eq. (\ref{start}) in the normal layer. The
latter has the form
\begin{eqnarray}
\left( \begin{array}{c} G_{\omega_n} (x,x')\\F^+_{\omega_n}
(x,x')\end{array}\right)=\left(
\begin{array}{c}-\frac{i}{v_x} e^{[ik_x-(\omega_n/v_x)]|x-x'|}\\0\end{array}
\right)+e^{-\omega_nx/v_x}e^{ik_x x} h(x')\left(
\begin{array}{c} 1\\0\end{array}\right)+ e^{\omega_nx/v_x}e^{-ik_x x} j(x')
\left(
\begin{array}{c} 1\\0\end{array}\right) \label{norm} \\
+e^{\omega_nx/v_x}e^{ik_x x} k(x')\left(
\begin{array}{c} 0\\1\end{array}\right)+e^{-\omega_nx/v_x}e^{-ik_x x} l(x')
\left(
\begin{array}{c} 0\\1\end{array}\right)\nonumber
\end{eqnarray}
and contain four additional unknown functions $h(x')$, $j(x')$,
$k(x')$ and $l(x')$. The boundary conditions at two $NS$ interfaces
provide eight equations which allow to uniquely determine all the
above functions and, hence, the supercurrent in $SINI'S$ junctions.
These equations are specified below.

Consider the left boundary. Making use of eq. (\ref{leftb1}) we find
\begin{equation}
\left( \begin{array}{c} h(x')e^{\frac{\omega_n d}{2v_x}}\\
k(x')e^{-\frac{\omega_n d}{2v_x}} \end{array}\right) =\alpha_1
\left(\begin{array}{c} 1\\-i e^{i\chi/2}\gamma^{-1}
\end{array}\right)e^{-\kappa d/2}f(x')+\beta_1
\left( \begin{array}{c} 1\\i e^{i\chi/2}\gamma
\end{array}\right)e^{-\kappa d/2}g(x'), \label{f}
\end{equation}
while eq. (\ref{leftb2}) yields
\begin{equation}
\left( \begin{array}{c} -\frac{i}{v_x}e^{ik_x x'} e^{-\frac{\omega_n}{v_x}x'}
e^{\frac{-\omega_n d}{2v_x}}\\ 0
\end{array}\right) +\left(
\begin{array}{c} j(x')e^{\frac{-\omega_n d}{2v_x}}\\ l(x')e^{\frac{\omega_n
d}{2v_x}}
\end{array}\right) =\alpha_1^* \left(\begin{array}{c} 1\\i e^{i\chi/2}\gamma
\end{array}\right)e^{-\kappa d/2}g(x')+\beta_1^*
\left( \begin{array}{c} 1\\-i e^{i\chi/2}\gamma^{-1}
\end{array}\right)e^{-\kappa d/2}f(x'). \label{s}
\end{equation}
Similarly, applying the boundary conditions at the right interface one gets
\begin{equation}
\left( \begin{array}{c} 1\\i e^{-i\chi/2}\gamma
\end{array}\right)e^{-\kappa d/2} n(x')=\alpha_2
\left( \begin{array}{c} -\frac{i}{v_x}e^{-ik_x x'} e^{\frac{\omega_n}{v_x}x'}
e^{\frac{-\omega_n d}{2v_x}}+e^{\frac{-\omega_n d}{2v_x}}h(x')\\
e^{\frac{\omega_n d}{2v_x}}k(x') \end{array}\right)+\beta_2 \left(
\begin{array}{c} e^{\frac{\omega_n d}{2v_x}}j(x')\\
e^{-\frac{\omega_n d}{2v_x}}l(x') \end{array}\right), \label{t}
\end{equation}
and
\begin{equation}
\left( \begin{array}{c} 1\\-i e^{-i\chi/2}\gamma^{-1}
\end{array}\right)e^{-\kappa d/2} r(x')=
\alpha_2^* \left(
\begin{array}{c} e^{\frac{\omega_n d}{2v_x}}j(x')\\
e^{-\frac{\omega_n d}{2v_x}}l(x') \end{array}\right)+ \beta_2^* \left(
\begin{array}{c} -\frac{i}{v_x}e^{-ik_x x'} e^{\frac{\omega_n}{v_x}x'}
e^{\frac{-\omega_n d}{2v_x}}+e^{\frac{-\omega_n d}{2v_x}}h(x')\\
e^{\frac{\omega_n d}{2v_x}}k(x') \end{array}\right). \label{fo}
\end{equation}
It is easy to see that the free terms in eqs. (\ref{f})-(\ref{fo})
are
\begin{equation}
z_1(x')=-\frac{i}{v_x}e^{ik_x x'-(\omega_n x'/v_x)},\:
z_2(x')=-\frac{i}{v_x}e^{-ik_x x'+(\omega_n x'/v_x)}.
\end{equation}
Eqs. (\ref{f})-(\ref{fo}) can be trivially resolved and we arrive at the
solutions for the functions $h(x')$ and $j(x')$ (which are only
needed in order to evaluate the current) with the structure
\begin{equation}
h(x')=U_1z_1(x')+U_2z_2(x'),\quad j(x')=V_1z_1(x')+V_2z_2(x'),\label{str}
\end{equation}
where $U_{1,2}$ and $V_{1,2}$ do not depend on $x'$.

\begin{multicols}{2}

\section{}
Consider the Eilenberger quasiclassical propagator which has a $2\times
2$ matrix structure in the Nambu space
\begin{equation} \hat g(\bbox{p},\bbox{R},\omega_n)=\left(
\begin{array}{cc} g(\bbox{p},\bbox{R},\omega_n) &
if(\bbox{p},\bbox{R},\omega_n)\\ -if^+(\bbox{p},\bbox{R},\omega_n) &
-g(\bbox{p},\bbox{R},\omega_n) \end{array} \right). \end{equation}
Here $\bbox{p}$ is the electron momentum on the Fermi surface
and $\bbox{R}$ is its coordinate.
This quasiclassical propagator obeys the normalization condition
$\hat g^2=1$ or, equivalently, $g^2+ff^+=1$. In addition, anomalous ($f,
f^+$) and normal ($g$) Green functions obey important symmetry relations
\begin{eqnarray}
&f^{+*}(\bbox{p},\bbox{R},\omega_n)=f(-\bbox{p},\bbox{R},\omega_n)=
f(\bbox{p},\bbox{R},-\omega_n), & \label{sym} \\ &
g^*(\bbox{p},\bbox{R},\omega_n)=g(-\bbox{p},\bbox{R},\omega_n)=
-g(\bbox{p},\bbox{R},-\omega_n).& \nonumber
\end{eqnarray}
The Eilenberger equations \cite{Eil} can be written in a concise matrix form as
\begin{eqnarray}
&i\bbox{v}_F\nabla \hat g+\hat\omega \hat g -\hat g\hat\omega=0,& \label{E}\\
&\displaystyle\hat \omega
=\bigl(i\omega_n+\frac{e}{c}\bbox{v}_F\bbox{A}\bigr)\hat\sigma_z
-\hat\Delta+\frac{i}{2\tau}\langle \hat g\rangle + \frac{i}{2\tau_s}\langle
\hat \sigma_z\hat g \hat\sigma_z\rangle,& \nonumber
\end{eqnarray}
where $\bbox{A}$ stands for the vector-potential; $\tau$ and $\tau_s$ are the
elastic scattering time on nonmagnetic and paramagnetic impurities,
respectively. The angular brackets denote averaging over the Fermi surface.
The matrix $\hat\Delta (\bbox{R})$ incorporates the superconducting order
parameter $\Delta (\bbox{R})$, $\hat\sigma_z$ is the Pauli matrix
\begin{equation} \hat\Delta=\left( \begin{array}{cc}0 & \Delta\\-\Delta^* & 0
\end{array}\right), \quad \hat\sigma_z=\left(
\begin{array}{cc} 1 & 0\\ 0 & -1 \end{array}\right).  \end{equation}
The current density $\bbox{j}(\bbox{R})$ is defined as follows
\begin{equation}
\bbox{j}(\bbox{R})= -2\pi i e TN(0)\sum_m\langle
\bbox{v}_F(\bbox{p})g(\bbox{p},\bbox{R},\omega_n)\rangle
\label{Jeil}
\end{equation}
$N(0)$ is the density of states at the Fermi energy per one spin direction.

The Eilenberger equations (\ref{E}) should be supplemented by Zaitsev boundary
conditions at metallic interfaces. These conditions have the form\cite{zait}
\begin{equation}
\hat g_{a+}=\hat g_{a-}=\hat g_a,
\label{bound1}
\end{equation}
$$\hat
g_a\bigl[(1-D(\bbox{p}))(\hat g_{s+}+\hat g_{s-})^2+ (\hat g_{s+}-\hat
g_{s-})^2\bigr]
$$
\begin{equation}
=D(\bbox{p})\bigl[ \hat g_{s+}\hat g_{s-}-\hat g_{s-}\hat
g_{s+}\bigr].
\label{bound2}
\end{equation}
Here $\hat g_{s,a}(\bbox{p},\bbox{R},\omega_n)= [\hat
g(\bbox{p},\bbox{R},\omega_n)\pm \hat g(\bbox{p}_r,\bbox{R},\omega_n)] /2$, by
$\bbox{p}_r$ we denote the reflected momentum. The subscripts $\pm$ in
eqs. (\ref{bound1}), (\ref{bound2}) stand for the expressions on the right (left)
side of the interface, respectively. Finally, $D(\bbox{p})$ is the
transparency coefficient of the boundary for the electron at the Fermi surface
with the given direction of momentum.  Eq. (\ref{bound1})
results in the current conservation at the boundary.

Consider the Josephson current in a clean $SINI'S$ structure in
the absence of the magnetic field. We assume that both $NS$ interfaces
are specularly reflecting and are perpendicular to the
$x$-axis. In this case the quasiclassical propagator depends on $p_x, x$ and
$\omega_n$. Making use of the symmetry relations (\ref{sym}) the functions
$g_{s,a}(p_x,x,\omega_n)=[g(p_x,x,\omega_n)\pm g(-p_x,x,\omega_n)]/2$ and
similarly defined functions $f_{s,a},f^+_{s,a}$ can be parametrized as follows
\begin{eqnarray}
&g_s=b_{s1},\: f_s=b_{s2}-ib_{s3},\: f^+_{s}=b_{s2}+ib_{s3},& \label{b}\\
&g_a=ib_{a1},\: f_a=b_{a3}+ib_{a2},\: f^+_{a}=-b_{a3}+ib_{a2},& \nonumber
\end{eqnarray}
The parameters $b_s, b_a$ are real and constitute two three-dimensional
vectors
$\bbox{b}_{s(a)}=(b_{s1(a1)},b_{s2(a2)},b_{s3(a3)})$. Combining
the Eilenberger equations for $\hat g(\pm p_x,
x,\omega_n)$, one easily finds
\begin{equation} \frac{d \bbox{b}_s}{d x}=\bbox{M}{\bf
\times}\bbox{b_a}, \quad \frac{d \bbox{b}_a}{d x}=-\bbox{M}{\bf
\times}\bbox{b_s}. \label{main} \end{equation}
Eqs. (\ref{main}) should be considered only for positive $p_x>0$ and $\omega_n>0$.
The three-dimensional vector $\bbox{M}$ in (\ref{main}) is real and
has the following components $M_1=2\omega/v_x$,
$M_2=(\Delta+\Delta^*)/v_x$ and
$M_3=i(\Delta-\Delta^*)/v_x$. Note that by introducing a complex
vector $\bbox{z}=\bbox{b}_s+i\bbox{b}_a$ one can rewrite
eqs. (\ref{main}) as $d
\bbox{z}/dx=-i\bbox{M}\times \bbox{z}$. From the latter
equation we conclude that $ \bbox{z}^2$ should be equal to a
constant. From the normalization condition one finds $
\bbox{z}^2=1$, or
\begin{equation}
\bbox{b}_s^2=1+\bbox{b}_a^2,\quad \bbox{b_s}\bbox{b_a}=0. \label{norm2}
\end{equation}
The boundary conditions (\ref{bound1}), (\ref{bound2}) take the form
\begin{eqnarray}
&&\bbox{b}_{a-}= \bbox{b}_{a+}=\bbox{b}_a, \label{zai2}\\
&&\bigl[(\bbox{b}_{s+}-\bbox{b}_{s-})^2+
(1-D)(\bbox{b}_{s+}+\bbox{b}_{s-})^2\bigr]\bbox{b}_a=2D \bbox{b}_{s+} \times
\bbox{b}_{s-} \nonumber
\end{eqnarray}

Assuming the pairing potential to be constant in the superconductors one
easily recover the solution of the Eilenberger equations. For the left
superconductor $x<-d/2$  we obtain
\begin{equation} \bbox{b}_s=\bbox{e}_{M-}+\bbox{C_-}\exp(|\bbox M|x),\quad
\bbox{b}_a = \bbox{b}_s \times \bbox{e}_{M-}. \label{ed}
\end{equation}
Here $\bbox{C_-}$ is an arbitrary vector perpendicular to $\bbox{M}$ and
$\bbox{e}_{M-}$ is the unit vector in the direction of $\bbox{M}$
\begin{equation}
\bbox{e}_{M-}=\frac{1}{\sqrt{|\Delta|^2+\omega_n^2}}\left( \begin{array}{c}
\omega_n\\ \Delta\cos(\chi/2)\\ \Delta\sin(\chi/2)
\end{array}\right). \label{e}
\end{equation}
Analogously, for $x>d/2$ we have
\begin{equation} \bbox{b}_s=\bbox{e}_{M+}+\bbox{C_+}\exp(-|\bbox M|x),\quad
\bbox{b}_a = -\bbox{b}_s \times \bbox{e}_{M+},
\end{equation}
where vector $\bbox{e}_{M+}$ is given by eq.(\ref{e}) with the changed sign of
$\chi$. Using these equations, from eq. (\ref{zai2}) one can establish the
relation between $\bbox{b}_a$ and $\bbox{b}_s$ at the normal side of the
interface near the left superconductor
\begin{equation}
\bbox{b}_a =t_1\bbox{b}_s\times\bbox{e}_{M-}.\label{bbb}
\end{equation}
Similarly, for the right boundary we get
\begin{equation}
\bbox{b}_a =t_2 \bbox{e}_{M+}\times \bbox{b}_s.\label{bbd}
\end{equation}
With the aid of these conditions one can easily find the Josephson current in
$SINI'S$ junctions. What remains is to solve the Eilenberger equations
(\ref{main}) in the normal metal. In the absence of external fields and
impurities we have
\begin{equation}
\bbox{b}_s=\left( \begin{array}{c} C\\ L_+\cosh\tilde x+L_-\sinh\tilde x\\
M_+\cosh\tilde x+M_-\sinh\tilde x \end{array}\right),
\end{equation}
\begin{equation}
\bbox{b}_a=\left(
\begin{array}{c} D\\ M_+\sinh\tilde x+M_-\cosh\tilde x\\ -L_+\sinh\tilde
x-L_-\cosh\tilde x \end{array}\right),
\end{equation}
where $\tilde x=2\omega_n x/v_x$ and $C,D, L_\pm, M_\pm$ are constants
determined from the normalization and boundary conditions. Finally, making use
of eq. (\ref{Jeil}) we arrive at the result (\ref{JN}),(\ref{Q}).

\end{multicols}
\end{document}